\documentclass[aip,jcp,reprint,twocolumn,a4paper,amsmath,floatfix]{revtex4-2}
\usepackage[utf8]{inputenc}
\usepackage{graphicx}
\usepackage{amsmath}
\usepackage{amssymb}
\usepackage{bm}
\usepackage{color}
\usepackage{hyperref}

\usepackage{printlen}

\begin{document}
\title{Freezing line of polydisperse hard spheres via direct-coexistence simulations}
\author{Antoine Castagn\`ede}
\email[\textbf{Author to whom correspondence should be addressed: }]{antoine.castagnede@universite-paris-saclay.fr}
\affiliation{Universit\'e Paris-Saclay, CNRS, Laboratoire de Physique des Solides, 91405 Orsay, France \looseness=-1}
\author{Laura Filion}
\affiliation{Soft Condensed Matter and Biophysics, Debye Institute for Nanomaterials Science, Utrecht University, Utrecht, Netherlands}
\author{Frank Smallenburg}
\affiliation{Universit\'e Paris-Saclay, CNRS, Laboratoire de Physique des Solides, 91405 Orsay, France \looseness=-1}

\begin{abstract}
    In experimental systems, colloidal particles are virtually always at least somewhat polydisperse, which can have profound effects on their ability to crystallize. Unfortunately, accurately predicting the effects of polydispersity on phase behavior using computer simulations remains a challenging task. As a result, our understanding of the equilibrium phase behavior of even the simplest colloidal model system, hard spheres, remains limited. Here, we present a new approach to map out the freezing line of polydisperse systems that draws on direct-coexistence simulations in the semi-grand canonical ensemble. We use this new method to map out the conditions where a hard-sphere fluid with a Gaussian size distribution becomes metastable with respect to partial crystallization into a face-centered-cubic crystal.  
    Consistent with past predictions, we find that as the polydispersity of the fluid increases, the coexisting crystal becomes increasingly size-selective, exhibiting a lower polydispersity and larger mean particle size than the fluid phase. Finally, we exploit our direct-coexistence simulations to examine the characteristics of the fluid-crystal interface, including the surface stress and interfacial absorption.
\end{abstract}
\maketitle

\section{Introduction}
\label{section:introduction}

When studying colloidal self-assembly, we often assume that we are working with systems that are either monodisperse, or consist of a limited number of distinct species, each with a well-defined size, shape, and other characteristics. In practice, however, colloidal particles inevitably exhibit a finite degree of polydispersity: variations of the particle characteristics around their mean value. This is particularly crucial in crystallization studies, where polydispersity can hinder the formation of well-ordered structures, or even induce formation of complex crystal phases \cite{cabane2016hiding, bommineni2019complex,coslovich2018local, lindquist2018communication}.

Unfortunately, we still have a limited understanding of the effect of polydispersity on the phase behavior of even the simplest of colloidal model systems: hard spheres. Colloidal hard spheres are arguably one of the most fundamental model systems in soft-matter physics, as their simplicity makes them extremely amenable for study in theory, simulations, and experiments \cite{royall2024colloidal}. As a result, significant efforts have been made to explore the effects of size polydispersity on e.g. the dynamics \cite{sear2000molecular, orsi2012dynamics, roberts2020dynamics}, phase behavior \cite{coslovich2018local, fasolo2003equilibrium, fasolo2004fractionation, bolhuis1996monte,  kofke1999freezing, wilding2010phase, bolhuis1996numerical, sollich2010crystalline} and crystallization \cite{auer2004quantitative, schope2006small, zaccarelli2009crystallization, bommineni2020spontaneous, lindquist2018communication, coslovich2018local, martin2003crystallization, lacour2022tuning, evans2001diffusive, kurita2012measuring} of hard spheres, as well as the behavior of polydisperse hard-sphere mixtures near a wall \cite{buzzacchi2004polydisperse, pagonabarraga2000local}. Despite these efforts, our knowledge of the equilibrium crystallization behavior is still incomplete. Specifically, simulation studies that focus on drawing equilibrium phase boundaries for polydisperse hard-sphere mixtures \cite{bolhuis1996monte, bolhuis1996numerical, kofke1999freezing, wilding2010phase, fasolo2003equilibrium, fasolo2004fractionation, sollich2010crystalline} typically consider only crystallization into face-centered cubic (FCC) crystals -- the crystal phase that is known to be stable for monodisperse hard spheres. However, a number of recent simulation studies have revealed that at relatively high polydispersities ($\gtrsim 10\%$) more complex crystal structures can spontaneously form, including Laves phases \cite{lindquist2018communication, coslovich2018local, bommineni2019complex}, whose stability has not yet been explored in detail. 

The reason for our lack of clear results for this fundamental model system is the fact that determining phase boundaries for polydisperse systems is  notoriously difficult \cite{sollich2010crystalline}. In polydisperse systems coexisting phases can have significantly different distributions of the relevant particle properties, a phenomenon known as fractionation. As a result, determining phase boundaries requires considering a vast space of possible mixtures of particles with different size distributions, significantly complicating any computational methodology.

Past studies have addressed this problem in a variety of ways, often making use of semi-grand canonical ensemble simulations, where the particles can fluctuate in size but the total number of particles stays fixed \cite{kofke1987nearly, bolhuis1996monte,  bolhuis1996numerical, wilding2010phase}. Using this ensemble, methodologies such as Gibbs-Duhem integration \cite{bolhuis1996monte, kofke1999freezing, bolhuis1996numerical} or phase-switch Monte Carlo \cite{wilding2000freezing, wilding2009solid, wilding2010phase} provide a route towards locating coexisting state points. Alternatively, theoretical methods can be used to estimate phase boundaries based on free-energy calculations that build on approximate equations of state for the fluid and crystal phases \cite{fasolo2003equilibrium, fasolo2004fractionation}.
These methods have given us a solid picture of the phase behavior of hard spheres in the low-polydispersity regime, but often struggle at higher polydispersities.

Here, we propose a new approach for predicting fluid-solid phase coexistences based on direct-coexistence simulations in the semi-grand canonical ensemble. In particular, we build on our recent work on direct-coexistence simulations for monodisperse systems \cite{smallenburg2024simple}, and combine it with efficient event-driven simulations in which particles can fluctuate in size based on an externally imposed relative chemical potential. Using this approach, we map out the fluid-FCC freezing line (i.e. the fluid cloud curve) as a function of polydispersity, and examine the behavior of the particle size distribution across the interface. Our methodology can be readily extended to other model systems and more complex crystal phases, opening the door to a more complete exploration of polydisperse phase behavior in the future.

\section{Methods}
\label{section:methods}

The main aim of this work is to determine the thermodynamic conditions for the fluid-solid phase coexistence of a polydisperse hard-sphere system, where the particle size distribution in the fluid phase is characterized by a Gaussian distribution
\begin{equation}
    \mathcal{P}(\sigma) = 
    \frac{1}{p \bar{\sigma} \sqrt{2 \pi}} \exp \left( - \frac{1}{2} \left( \frac{\sigma - \bar{\sigma}}{p \bar{\sigma}} \right)^2 \right).
\end{equation}
Here, $p$ is the polydispersity of the parent fluid phase, defined as the ratio of the standard deviation of the particle diameter $\sigma$ to its mean value $\bar{\sigma}$. 
Note that fixing the size distribution of the fluid phase does not fix the overall size distribution in the coexisting system. Since the crystal phase is generally expected to have a different size distribution (and polydispersity) from the fluid, the overall size distribution of any coexisting state also depends on the fraction of volume taken up by each phase. In this work, we specifically focus on the identification of the freezing line of a fluid with fixed size distribution, also known as the fluid \textit{cloud curve} \cite{fasolo2003equilibrium}. The corresponding set of coexisting crystal phases is known as the \textit{shadow curve}.


When two polydisperse phases coexist, thermodynamic equilibrium requires that their temperature $T$, pressure $P$, and \textit{absolute} chemical potential $\mu(\sigma)$ are equal in the two phases. The core idea of the strategy we propose here is to look at a series of ``trial'' fluids at different densities, and to check whether there is a polydisperse crystal phase which satisfies these criteria. To narrow down the search for this crystal phase, we note that a weaker (and insufficient) requirement for such a phase is that it has the same temperature, pressure, and \textit{relative} chemical potential $\delta \mu(\sigma)$. Here, $\delta \mu(\sigma) = \mu(\sigma) - \mu(\sigma_\mathrm{ref})$ represents the difference in chemical potential between particles of size $\sigma$ and particles of an arbitrary reference size $\sigma_\mathrm{ref}$. 
Since the combination of $T$, $P$ and $\delta \mu(\sigma)$ uniquely define a crystal state point, there is only one trial crystal state that can coexist with a given trial fluid. 
The final requirement for an equilibrium coexistence is then the constraint that the \textit{absolute} chemical potentials of the two phases are also equal.
This can be checked in a direct-coexistence simulation by confirming that the trial crystal can coexist with the initial trial fluid without deforming.
For this, we use the trick of Ref. \onlinecite{smallenburg2024simple} of checking for crystal deformation by monitoring the pressure component $P_{zz}$ along the direction perpendicular to the interface between the two phases.

Finding the equilibrium freezing point for a polydisperse system with a given polydispersity $p$ and temperature $T$ therefore comes down to the following steps:
\begin{itemize}
    \item Choose a series of ``trial'' fluid densities $\rho_f^\mathrm{trial}$, and measure the corresponding pressures $P_f^\mathrm{trial}$ and relative chemical potentials $\delta \mu_f^\mathrm{trial}(\sigma)$.
    \item For each $\rho_f^\mathrm{trial}$, determine the density (and corresponding lattice spacing) of the corresponding trial crystal phase that could coexist with this fluid.
    In other words, find the equilibrium bulk crystal that is stable at $P_f^\mathrm{trial}$ and $\delta\mu_f^\mathrm{trial}(\sigma)$. 
    \item For each $\rho_f^\mathrm{trial}$, perform a direct-coexistence simulation in the semi-grand canonical ensemble between the trial fluid and trial crystal phase.
    If the coexistence remains stable, check for deformation of the crystal phase by measuring $P_{zz}$. 
\end{itemize}
After these steps, the identification of the equilibrium freezing line is done by determining the value of $\rho_f^\mathrm{trial}$ where $P_{zz}$ in the direct-coexistence simulation is equal to $P_f^\mathrm{trial}$. Note that the above steps ensure that the fluid phase in the direct coexistence simulation has the same pressure and relative chemical potential as the trial fluid, which fixes its size distribution to be equal to that of the trial fluid as well.

In the following subsections, we outline each of these steps in more detail.

\subsection{Chemical potential of the fluid phase}
\label{subsection:chemeq}

We start by measuring the bulk properties of our trial fluids. To this end, we perform event-driven molecular dynamics (EDMD) simulations (based on Ref. \onlinecite{smallenburg2022efficient}) of polydisperse hard-sphere systems in the canonical ensemble. We use system sizes of both 16000 and 32000 particles, and simulate for a total simulation time of $1.5\times 10^5 \tau$ ($1/3$ of it being dedicated to equilibration).
Here, our time unit is $\tau = \sqrt{\beta m \sigma^2}$, where $\beta = 1 / k_B T$ and $m$ is the particle mass, which is taken to be the same for all particles. Hence, $\tau$ corresponds to the time it takes a typical particle in free flight to move over a distance roughly equivalent to its own diameter. For the sake of reproducibility and to avoid significant deviations from the expected size distribution, we chose to generate particle sizes deterministically using the inverse of the cumulative probability distribution function.
Further details on this can be found in the supplementary material (SM).

From each of these simulations, our aim is to determine both the pressure $P$ and the relative chemical potential $\delta \mu (\sigma)$.
We obtain the pressure via the standard virial expression for hard-sphere systems
\cite{alder1960studies}:
\begin{equation}\label{eq:pressure}
    \beta P_{kl}/\rho = 1 +  
    \frac{\sum\limits_c m {v}_{ij}^{k} {r}_{ij}^{l}}{ N (t_b - t_a)},
\end{equation}
where $P_{kl}$ indicates the $kl$-component of the pressure tensor (with $k,l \in \{x,y,z\})$, $\rho=N/V$ is the number density, and the sum is taken over all  collisions in a time interval $[t_a ; t_b]$. Additionally, 
${r}_{ij}^{k}$ and ${v}_{ij}^{k}$ are the $k$-components of the relative distance and velocity vectors between the colliding particles $i$ and $j$, respectively.

For polydisperse mixtures, the total chemical potential can be expressed as the sum of two contributions, namely the mixing part $\mu_\textrm{mix}(\sigma)$ and the configurational part $\mu_\textrm{conf}(\sigma)$, which are both functions of the particle size.
The former is an intrinsic property of the particle size distribution, given by $\beta \mu_\textrm{mix}(\sigma) = -\log\left(\mathcal{P}(\sigma)\lambda\right)$, with $\lambda$ an arbitrary length scale analogous to the thermal wavelength.
The derivative of the latter, $\mu'_\textrm{conf}(\sigma)$, can be linked to the internal energy of the system following (see SM for details)
\begin{equation}
    \label{eq:muconf}
    \mu'_\textrm{conf}(\sigma_i) = 
    \left\langle \frac{\partial U}{\partial \sigma_i} \right\rangle,
\end{equation}
where $\left\langle~\dots~\right\rangle$ denotes an ensemble average, $U$ is the potential energy of the system, and $\sigma_i$ refers to the size of particle $i$.

As we show in the SM, Eq. \ref{eq:muconf} can then be interpreted in terms of an average force experienced by the particle radius $R_i$: 
\begin{equation}
    \label{eq:muconf2}
    \mu'_\textrm{conf}(\sigma_i) = - \langle f_{R_i}\rangle / 2,
\end{equation}
where a positive $f_{R_i}$ corresponds to an outward force on the particle surface.
In practice, for hard spheres, the forces $\mathbf{f}_i$ on any particle's surface are due to collisions, and always directed radially inwards, such that $f_{R_i}$ is always negative (and hence $\mu'_\textrm{conf} > 0$). 
We can write the ensemble average:
\begin{equation}
    \langle f_{R_i} \rangle =  \frac{-1}{t_b - t_a} \sum_k m \, |(\delta \mathbf{v}_i)_k|,
\end{equation}
where $(\delta \mathbf{v}_i)_k$ is the change in velocity of particle $i$ during collision $k$, and the sum is taken over all collisions in the time interval $[t_a, t_b]$. 

Combining $\mu'_\textrm{conf}(\sigma)$ and $\mu'_\textrm{mix}(\sigma)$, we can calculate the derivative of the total chemical potential, $\mu'(\sigma)$, which can be fitted and integrated to obtain $\delta \mu(\sigma)$ -- the function needed to perform semi-grand canonical simulations.

\begin{figure*}
    \centering
    \includegraphics[width=\textwidth]{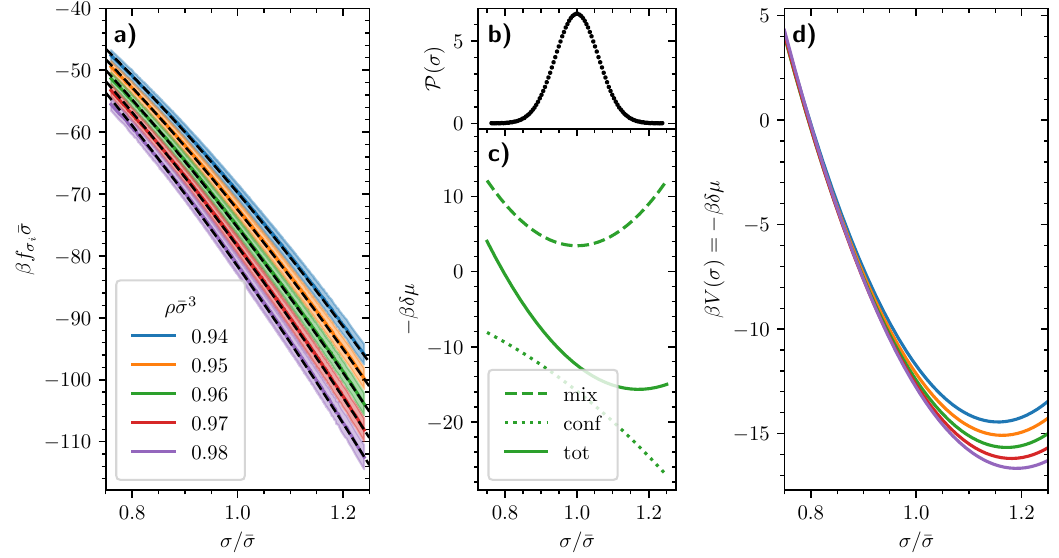}
    \caption{
        a) Surface forces $\mathbf{f}_i = - f_{\sigma_i}$ measured for a 16000 particle system with Gaussian particle size distribution and polydispersity $p = 0.06$, for a range of system densities $\rho \bar{\sigma}^3$ (top to bottom: 0.94, 0.95, 0.96, 0.97, and 0.98).
        Shading corresponds to one standard error and dashed black lines represent the fits $f\left(\sigma, \rho\bar{\sigma}^3\right)$ we employed in subsequent simulations.
        b) Particle size distribution for a system with polydispersity $p=0.06$.
        c) Configurational contribution (dashed line), mixing contribution (dotted line), and total chemical potential difference for the above system at density 0.96.
        d) Total chemical potential difference for the systems shown in the left panel.
    }
    \label{fig:forceschempot}
\end{figure*}

The typical behavior of the surface forces is shown in Fig. \ref{fig:forceschempot}a for systems with polydispersity $p = 0.06$ (see particle size distribution in Fig. \ref{fig:forceschempot}b). We observe that the forces become monotonically stronger as the density increases -- which makes sense when considering that higher densities will lead to a higher collision rate. Similarly, the forces are stronger for larger particle sizes, which can be understood from the fact that larger particles will experience more collisions on their larger surface. For all of our data, a quadratic fit in $\sigma$ is in good agreement with the measured forces.
To obtain a consistent fit across all investigated densities, we fit the forces with a polynomial function of the form
\begin{equation}
    \label{eq:forcesfit}
   \beta f(\sigma, \rho\bar{\sigma}^3) \bar{\sigma} = \sum_{i=0}^{2} \sum_{j=0}^{2} c_{i,j} \left(\rho \bar\sigma^3\right)^i
   \left(\frac{\sigma}{\bar{\sigma}}\right)^j.
\end{equation}
The outcome of this fit $f\left(\sigma, \rho\bar{\sigma}^3\right)$ is shown in dashed lines in Fig. \ref{fig:forceschempot}a, and is in good agreement with the computed forces.
This procedure allows us to explore systems with any density in this range without having to recalculate the forces.
We have confirmed that surface forces measured in larger systems (up to 32000 particles) are consistent with our predictions in order to rule out significant finite-size effects.

We show the different contributions to the chemical potential difference $\delta \mu \left(\sigma, \rho\bar{\sigma}^3\right)$ in Fig. \ref{fig:forceschempot}c for a system with polydispersity $p = 0.06$ and density $\rho\bar{\sigma}^3 = 0.96$.
Fig. \ref{fig:forceschempot}d shows the total chemical potential difference for a range of densities for this same system. Note that in Fig. \ref{fig:forceschempot}d we plot $-\delta \mu(\sigma)$ as this corresponds to the external potential field $V$ that we impose in the semi-grand-canonical simulations, described in the next section.

\subsection{Mechanical equilibrium with the crystal phase}
\label{subsection:mecheq}
Now that we have access to the behavior of the chemical potential difference as a function of system size, $\delta \mu\left(\sigma\right)$, we turn to simulations of hard spheres in the semi-grand canonical ensemble.
In a similar approach to the continuous time swap approach developed in Ref. \onlinecite{berthier2019efficient}, we consider a semi-grand ensemble simulation in which particles have three positional degrees of freedom supplemented by a fourth one: their radius $R_i = \sigma_i/2$. 
The Hamiltonian for this system can then be written as:
\begin{equation}
    \label{eq:hamiltonian}
    H = \sum_i \left[ \frac{\mathbf{p}_i^2}{2m} + \frac{\wp_i^2}{2M} + V(R_i)\right] + \sum_{i<j}U_{ij}(\mathbf{r}_i, \mathbf{r}_j, R_i,R_j),
\end{equation}
where $\mathbf{p}_i$ is the translational momentum of particle $i$, $m$ is the (translational) particle mass, $\wp_i$ is the momentum associated with its radius $R_i$, $M$ is the corresponding mass, $V(R_i) = -\delta\mu\left(\sigma_i=2R_i\right)$ is the external field controlling the particle size, and $U_{ij}$ represents the pair interaction (which is either 0 or $+\infty$ for hard spheres).

The Hamiltonian in Eq. \ref{eq:hamiltonian} naturally gives rise to equations of motion for the positions and radii of the particles.
Note that since our interaction potential is discontinuous, forces between the particles only occur at contact, and the collision rules can be derived by regarding the hard-sphere interaction as the limiting case of a sharply repulsive continuous interaction (see SM).
These collision rules can be straightforwardly implemented in an EDMD simulation.
However, the motion of the radial degree of freedom is additionally affected by the continuous field $V(R_i)$, which is not directly compatible with the event-driven nature of EDMD simulations.
Fortunately, this field  can be efficiently handled with the event-driven Monte Carlo (EDMC) approach introduced by Peters and De With \cite{peters2012rejection, castagnede2024fast}.
In this event-driven approach, continuous interactions (or in this case, fields) are handled stochastically.
Specifically, whenever we predict future collisions for a particle, we also consider collisions with the field $V(R_i)$, which are determined based on a maximum energy increase $\Delta V$ drawn from a Boltzmann distribution (see SM for details).
Collisions with the field $V(R_i)$ simply reverse the associated momentum $\wp_i$ for the particle under consideration.

\begin{figure}
    \centering
    \includegraphics[width=\columnwidth]{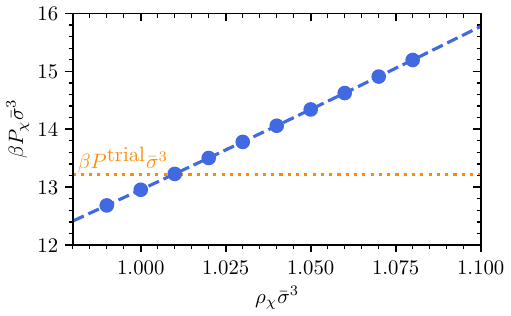}
    \caption{
        Crystal pressure as a function of crystal density obtained from semi-grand simulations of a bulk crystal of 16384 particles.
        The imposed chemical potential difference corresponds to that of a trial fluid system with polydispersity $p = 0.06$ at density $\rho_f^\textrm{trial} \bar{\sigma}^3 = 0.9618$.
        Pressure of the trial fluid is reported on the horizontal dotted line.
        Standard errors are shown for all points but are typically found to be smaller than the point size.
        The dashed line corresponds to the quadratic fit employed to determine the crystal density at mechanical equilibrium.
    }
    \label{fig:findmecheq}
\end{figure}

We now use the resulting semi-grand canonical simulations to identify the trial crystal phase that could coexist with a given trial fluid. 
Note that for a given $\rho_f^\textrm{trial}$, we  already know the pressure $P^\textrm{trial}$ and the associated chemical potential function $\delta \mu(\sigma)$.
The coexisting crystal phase must match the fluid in both of these quantities.
To find this coexisting crystal, we perform simulations of FCC crystals at different densities $\rho_\chi$ but at the same relative chemical potential $\delta\mu^\mathrm{trial}(\sigma)$, and measure the pressure in each simulation.
We then locate the density $\rho_\chi^\textrm{trial}$ such that $P_\chi\left(\rho_\chi^\textrm{trial}\right) = P^\textrm{trial}$.
Figure \ref{fig:findmecheq} shows this procedure for a system with polydispersity $p = 0.06$ and trial fluid density $\rho_f^\textrm{trial} \bar{\sigma}^3 = 0.9618$.
We used crystal bulk systems with 16384 particles set up on a FCC lattice, which were equilibrated for $25 \times 10^3 \tau$ and sampled over an additional $25 \times 10^3 \tau$.
We find that the pressure is a monotonically increasing function of $\rho_\chi$ (as expected), and simply locate the intersection of $P_\chi\left(\rho_\chi^\textrm{trial}\right)$ and $P^\textrm{trial}$ by fitting the former with a polynomial and numerically locating the crossing point based on the fit. 

\subsection{Semi-grand direct-coexistence simulations}
\label{subsection:SGDC}
\begin{figure*}
    \centering
    \includegraphics[width=\textwidth]{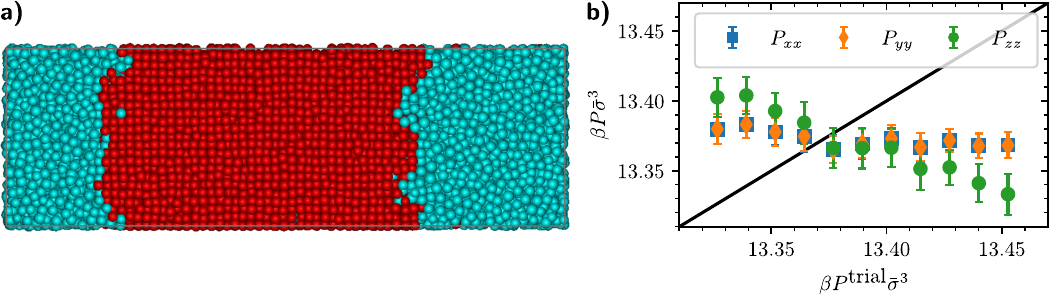}
    \caption{
        a) Snapshot of a fluid with polydispersity $p = 0.06$ in coexistence with a crystal.
        For visualization purposes, particles found to be in a local crystalline environment using ten Wolde's local bond order parameter\cite{wolde1996simulation} $q_6$ are displayed in red, while particles found to be fluid-like are displayed in blue.
        b) Pressure tensor components $P_{xx}$, $P_{yy}$, and $P_{zz}$ as a function of trial pressure $P^\textrm{trial}$ obtained from semi-grand direct-coexistence simulations for a 16384 particle system in which the fluid has polydispersity $p = 0.06$.
        Error bars show one standard error.
        These systems were constructed such that the fluid--crystal interface corresponds to the square-lattice plane of an FCC crystal.
    }
    \label{fig:dirco}
\end{figure*}

At this point, for each given trial fluid, we have identified a crystal phase that could coexist with it: the two phases have the same temperature, pressure, and relative chemical potential. However, we have not yet confirmed that this coexistence is indeed a stable equilibrium, as there is no guarantee that the trial fluid and the associated crystal have the same \textit{absolute} chemical potential.
To address this, we now perform semi-grand direct-coexistence simulations following the methodology developed in Ref. \onlinecite{smallenburg2024simple}.
We initialize a crystal slab with density $\rho_\chi^\textrm{trial}$ in an elongated box in the $z$ direction with aspect ratio $1:3$.
The global density inside the box is set to $(\rho_f^\textrm{trial} + \rho_\chi^\textrm{trial}) / 2$, and approximately half of it is initially filled with crystal.
The rest of the box is filled with smaller fluid particles to match the desired global density.
These particles are then rapidly grown to their target size distribution and the system is equilibrated for $10^5 \tau$.
A typical snapshot of a coexisting system is shown in Fig. \ref{fig:dirco}a.

We note that by using this choice for the global density, we place our coexisting system essentially at the midpoint of the tie-line that connects the cloud and shadow point corresponding to the two coexisting phases. This is similar in spirit to the approach used in Ref. \onlinecite{buzzacchi2006simulation} for determining the cloud point in a polydisperse lattice gas. Rather than direct coexistence simulations, Ref. \onlinecite{buzzacchi2006simulation} used grand-canonical ensemble simulations and iterative histogram reweighting approaches to locate coexistence where one of the two fluid phases has the desired size distribution. While this approach would be impractical for our system (e.g. because particle insertions are not effective in dense or crystalline phases), this has the similar effect of setting up a coexistence situation where each phase occurs with equal weight in order to locate the cloud point.

To determine whether the obtained coexistence is thermodynamically stable, we check whether the fluid and crystal phase remain at their initial state point by monitoring the pressure tensor. If the chemical potential is not equal in the two phases, their densities will adjust to reach thermodynamic equilibrium. This results in a deformation of the crystal along the long axis of the box (as the lattice spacings in the other directions are fixed by the box) and a change in overall density in the fluid.
This deformation will inevitably change the pressure component $P_{zz}$ (along the long axis of the box), which is homogeneous throughout the system \cite{smallenburg2024simple}.
In contrast, at equilibrium coexistence, both phases keep their initial density, and $P_{zz}$ will be equal to $P^\textrm{trial}$, resulting in an unstrained crystal coexisting with the trial fluid, at equal temperature, pressure, and chemical potential. Moreover, since both the pressure and relative chemical potential of the fluid phase in this coexistence simulation match those of our bulk trial fluid, its size distribution necessarily also matches that of the trial fluid. In contrast, the crystal phase will generally have a significantly different size distribution. Hence, the phase coexistence we found corresponds exactly to the fluid cloud point and the associated FCC crystal shadow phase.

Fig. \ref{fig:dirco}b shows the result of this procedure for the $p=0.06$ system, repeated for a range of trial fluids. Specifically, we plot the measured value of $P_{zz}$ (green points) as a function of the trial pressure $P^\mathrm{trial}$, and numerically locate the crossing point with the line $P_{zz} = P^\mathrm{trial}$. This yields the freezing point for this polydisperse mixture. Note that while significant noise is visible in Fig. \ref{fig:dirco}b, the effect of this noise on the resulting coexistence pressure is weak: we estimate the coexistence pressure to be $\beta P^\mathrm{coex}\bar{\sigma}^3 = 13.37(1)$.

Additionally, we plot the components of the pressure tangential to the interfaces ($P_{xx}$ and $P_{yy}$), which are related to the stress on the simulation box caused by the interfaces \cite{smallenburg2024simple}. As expected from the choice of crystal lattice orientation inside the box (square plane facing the fluid), we recover the equality between the $x$ and $y$ components of the pressure tensor.

\section{Results and discussion}
\label{section:results}

\begin{figure*}
    \centering
    \includegraphics[width=\textwidth]{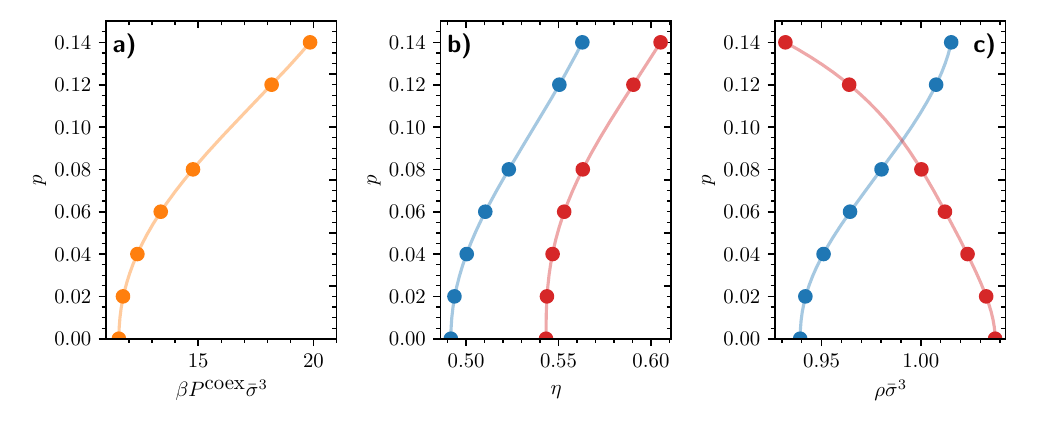}
    \caption{
        a) Coexistence pressure as a function of the polydispersity of the fluid phase.
        b) Freezing line (blue) and corresponding shadow curve (red) in the polydispersity-packing fraction plane.
        c) Freezing line and  shadow curve in the polydispersity-density plane.
        Data presented here corresponds to a fluid in coexistence with a FCC crystal, the fluid-crystal interface corresponds to the square lattice plane of the FCC crystal.
        Note that error bars are shown on all points but are smaller than the point size (typically on the order of $0.1\%$).
        All lines correspond to the functional forms reported in the main text.
        Data for the monodisperse case was obtained from Ref. \onlinecite{smallenburg2024simple}.
    }
    \label{fig:phasediag}
\end{figure*}

\begin{figure}
    \centering
    \includegraphics[width=\columnwidth]{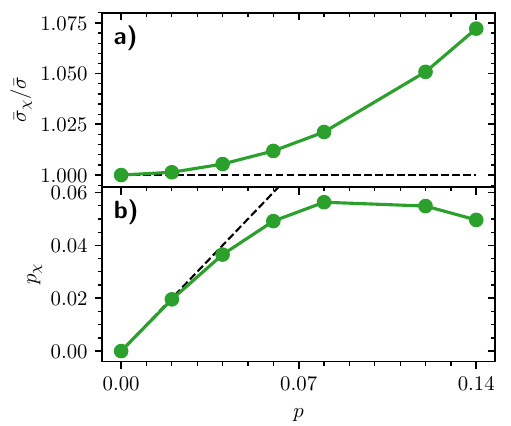}
    \caption{Mean particle size (a) and polydispersity (b) of the crystal phase as a function of the fluid polydispersity $p$. The dashed lines indicate the corresponding mean size and polydispersity for the fluid phase for comparison.}
    \label{fig:crystalcomposition}
\end{figure}

\begin{figure}
    \centering
    \includegraphics[width=\columnwidth]{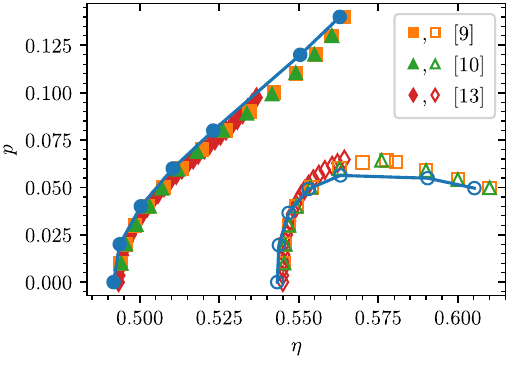}
    \caption{
        Fluid cloud curve and the associated crystal shadow curve in the packing fraction - polydispersity plane. Results from this work are shown as blue points and lines. We compare these to data points obtained from Refs. \onlinecite{fasolo2003equilibrium, fasolo2004fractionation, wilding2010phase}, which correspond respectively to a symmetric triangular, Schultz, and top-hat particle size distribution of the parent phase.
    }
    \label{fig:comparison}
\end{figure}

To study the full extent of the fluid-FCC freezing lines in hard-sphere mixtures with Gaussian polydispersity, we repeat the procedure outlined in the previous sections on systems with different polydispersities ranging from $2\%$ to $14\%$. 
The main results for the phase behavior are shown in Fig. \ref{fig:phasediag}. Specifically, in Fig. \ref{fig:phasediag}a we report the fluid-FCC coexistence pressure as a function of polydispersity. As expected, increasing polydispersity destabilizes the crystal, resulting in higher coexistence pressures. The same trend is seen in the packing fractions $\eta$ of the two coexisting phases (Fig. \ref{fig:phasediag}b), which both monotonically increase with increasing $p$. Intriguingly, plotting the number density $\rho = N/V$ of the two coexisting phases (Fig. \ref{fig:phasediag}c) shows a different perspective. While the number density of the coexisting fluid increases with polydispersity, the corresponding crystal density decreases. This is indicative of the fact that with increasing polydispersity, the mean size of the particles also increases, as shown in Fig. \ref{fig:crystalcomposition}a. This results in a larger lattice spacing and hence a lower number density. 
At a polydispersity close to $p \simeq 9\%$, the number densities of the fluid and crystal phase cross. This renders the precise measurement of the coexistence points more difficult. Specifically, when the density difference between the two phases is small, fluctuations in the amount of crystal phase in the direct-coexistence simulations grow larger, making it more likely that the crystal fully melts or starts percolating the box. Hence, in this regime we cannot keep the coexistence stable over long enough simulation times to accurately measure the pressure tensor. Due to this, we were unable to accurately determine the coexistence conditions at $p = 10\%$ and omit this point from Fig. \ref{fig:phasediag}.
At even higher polydispersities ($p \geq 12\%$) we can once again obtain a stable coexistence. In this regime, the number density in the FCC phase is lower than in the fluid, consistent with earlier predictions by Wilding and Sollich \cite{wilding2010phase}. 

Empirically, we find that the coexistence conditions are well-fitted by the following functions:
\begin{align}
    \beta P^\textrm{coex} \bar{\sigma}^3 &= 11.5646 + 433.63950 p^2 + 2308.37288 p^3\\
    \notag &~~~~- 19711.1098 p^4 + 19077.0257 p^5,\\
    \eta_f &= 0.4918 + 4.90278 p^2 + 23.48901 p^3\\
    \notag &~~~~- 375.76734 p^4 + 1021.79611 p^5,\\
    \eta_\chi &= 0.5432 + 0.31861 p^2 + 60.26721 p^3\\
    \notag &~~~~- 361.66200 p^4 +545.04336 p^5,\\  
    \rho_f \bar{\sigma}^3 &= 0.93918 + 5.63121 p^2 + 93.78531 p^3\\
    \notag &~~~~- 1645.24628 p^4 + 8878.27295 p^5\\
    \notag &~~~~- 18195.4842 p^6,\\
    \rho_\chi \bar{\sigma}^3 &= 1.0375 - 14.54067 p^2 + 184.78167 p^3\\
    \notag &~~~~- 1074.12989 p^4 + 1577.06258 p^5.
\end{align}
Here, $\eta_f$ and $\eta_\chi$ are the packing fractions of the fluid and FCC phase at coexistence, and $\rho_f$ and $\rho_\chi$ are the corresponding number densities.

Taking a closer look at the behavior of the composition of the crystal line along the shadow curve, we observe that while the mean particle size rises significantly with increasing fluid polydispersity, the polydispersity of the crystal saturates, approaching values around $5.6\%$ (Fig. \ref{fig:crystalcomposition}b), consistent with past predictions using various parent size distributions \cite{bolhuis1996monte, kofke1999freezing, fasolo2003equilibrium, fasolo2004fractionation}.

To compare to past predictions in more detail, we plot in Fig. \ref{fig:comparison} the fluid cloud curve and the corresponding FCC shadow curve in the packing fraction - polydispersity plane. We compare these curves to the results  from Refs. \onlinecite{fasolo2003equilibrium, fasolo2004fractionation, wilding2009solid}, which correspond respectively to hard-sphere fluids with a symmetric triangular, Schultz, and top-hat particle size distribution.
Note that in this representation, we use the polydispersity of the crystal phase for the crystal branch. Despite the differences in size distribution, we observe good agreement between all sets of data. Nonetheless, our new results deviate more from the data of of Refs. \onlinecite{fasolo2003equilibrium, fasolo2004fractionation, wilding2010phase} than the latter three sets differ between each other. In fact, the deviations already start in the limit of zero polydispersity, where the distribution shape should not matter. We attribute this deviation to inaccuracies in the equations of state used in Refs. \onlinecite{fasolo2003equilibrium, fasolo2004fractionation, wilding2010phase}. Specifically, for the fluid phase, these works use the  Boublik, Mansoori, Carnahan, Starling and
Leland (BMCSL) expression \cite{boublik1970hard, mansoori1971equilibrium}, which reduces to the standard Carnahan-Starling expression \cite{carnahan1969equation} in the monodisperse case. For monodisperse hard spheres, Carnahan-Starling is known to underestimate the pressure at high densities \cite{pieprzyk2019thermodynamic}. From our own pressure measurements (not shown), the BMCSL similarly underestimates the pressure of polydisperse mixtures. Such deviations would cause a shift in the freezing transition to higher packing fractions. 
This is consistent with the data of Refs. \onlinecite{fasolo2003equilibrium, fasolo2004fractionation, wilding2010phase}, who report the monodisperse freezing and melting packing fractions at approximately 0.494 and 0.545, respectively, which are high compared to modern estimates of approximately 0.492 and 0.543 \cite{royall2024colloidal, smallenburg2024simple}. 

\begin{figure}
    \centering
    \includegraphics[width=\columnwidth]{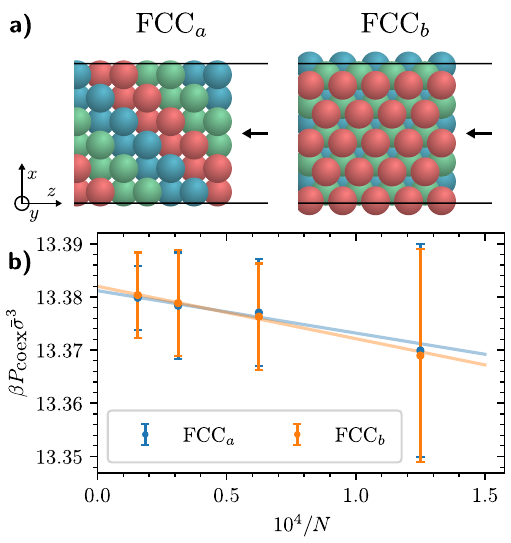}
    \caption{
        a) Schematic representation of the choice of crystal orientation in the simulation box. Note that the FCC$_b$ orientation used in this work differs from the one used in Ref. \onlinecite{smallenburg2024simple} by 90 degree rotation in the $xz$ plane.
        b) Coexistence pressure estimates as a function of system size for a system with polydispersity $p = 0.06$, and for the two different FCC crystal lattice orientations shown in the top panel.
        Lines are guides to the eye. Error estimates for $P^\textrm{coex}$ are determined from the noise in the individual pressure measurements (e.g. those shown in Fig. \ref{fig:dirco}).
    }
    \label{fig:sizedep}
\end{figure}

We expect the phase behavior reported in Fig. \ref{fig:phasediag} to be sensitive to finite-size effects which will typically serve to help stabilize the crystal phase. The data we report in Fig. \ref{fig:phasediag} is obtained for systems with $N=16000$ particles and the FCC lattice orientation in the box was chosen so that the square crystal plane is facing the fluid. To examine the strength of the finite-size effects, we repeat our calculations for different system sizes (up to $N=64000$) for both the same orientation of the crystal phase (denoted FCC$_a$) and one where the hexagonal planes are aligned along one of the long walls of the box (FCC$_b$, see Fig. \ref{fig:sizedep}a).
For small system sizes, we find variations in our predictions of the coexistence pressure for $p = 0.06$ as we report in Fig. \ref{fig:sizedep}b. Additional data on these different system sizes is provided in the SM.
Nevertheless, for the largest systems considered, coexistence pressure predictions are indistinguishable for the two crystal orientations we consider. This is consistent with the fact that the crystal orientation should not affect coexistence conditions in the thermodynamic limit.

We now turn to the characterization of the fluid-crystal interface at the freezing points determined above. 

\begin{figure}
    \centering
    \includegraphics[width=\columnwidth]{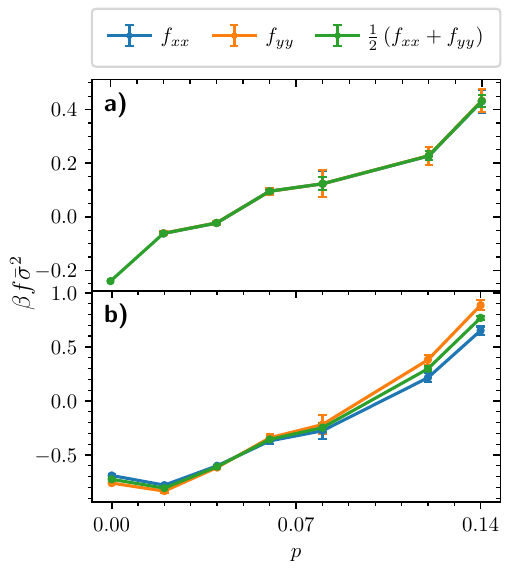}
    \caption{
        Surface stresses measured at coexistence for polydisperse systems in the case of a FCC$_a$ (a) and FCC$_b$ (b) fluid-crystal interface plane.
        Error bars correspond to twice the standard error obtained from 10 to 20 independent runs at the coexistence we determine for 16384 particles systems. Note that the actual error is likely to be larger, due to systematic errors introduced by inaccuracies in our determination of the coexistence point.
        Points for the monodisperse case were obtained from Ref. \onlinecite{de2024statistical} for FCC$_a$ and from additional monodisperse direct-coexistence simulations for FCC$_b$.
        All lines are guides to the eye.
    }
    \label{fig:surfacestress}
\end{figure}

First, we measure the surface stresses at the fluid-crystal interface. We can write the $xx$-component of the surface stress as \cite{de2024statistical}
\begin{equation}
    f_{xx} = \frac{1}{2} L_z \left(P_{zz} - P_{xx}\right),
\end{equation}
where $L_z$ denotes the length of the simulation box in the elongated $z$ direction. The expression for the $yy$ component is analogous.
The stress tensor is measured directly from semi-grand simulations at the coexistence points determined above, for systems of 16000 particles equilibrated for $10^5 \tau$.
These quantities are reported in Fig. \ref{fig:surfacestress}, along with $\frac{1}{2} (f_{xx} + f_{yy})$, which for the specific case of the square symmetry of the FCC$_a$ interface corresponds to the scalar value of the surface stress.

For the FCC$_a$ interface, increasing polydispersity rapidly causes the surface stress to depart from the known value for the monodisperse case \cite{davidchack1998simulation, de2024statistical}.
This is not the case for the FCC$_b$ interface, for which the surface stresses remain relatively constant in the low polydispersity regime, and close to the monodisperse values we evaluated from additional direct-coexistence simulations.
Upon increasing polydispersity, we notice an upward trend for both the FCC$_a$ and FCC$_b$ interfaces. For the two cases, the surface stress becomes positive at polydispersities around $0.06$ and $0.1$, respectively. 
These measurements are particularly sensitive to the precise determination of the coexistence point, which translates into the error bars reported in Fig. \ref{fig:surfacestress}.
While the sign change of the surface stress $f$ may at first glance appear surprising, one can recall the Shuttleworth equation\cite{shuttleworth1950surface, di2020shuttleworth} that links it to the surface free energy $\gamma$:
\begin{equation}
    f_{ij} = \delta_{ij} \gamma + \frac{\partial \gamma}{\partial u_{ij}},
    \label{eq:shuttle}
\end{equation}
with $u_{ij}$ the surface strain tensor and $\delta_{ij}$ the Kronecker delta. For monodisperse hard spheres, the surface stress of the investigated interfaces is negative. As $\gamma$ must be positive, this negative sign indicates that the second term in Eq. \ref{eq:shuttle} is negative: stretching the interface results in a lower $\gamma$. The increase in $f$ with increasing polydispersity can then be attributed to two possible effects. First, the rise in coexistence pressure with increasing polydispersity likely gives rise to higher surface free energies $\gamma$. Second, the ability of particle sizes to adapt to the structure of the interface may make $\gamma$ less dependent on the strain applied to the interface, which would lower the magnitude of the (negative) second term. A more careful analysis of the surface free energy would be required to disentangle these two effects.
\begin{figure*}
    \centering
    \includegraphics[width=\textwidth]{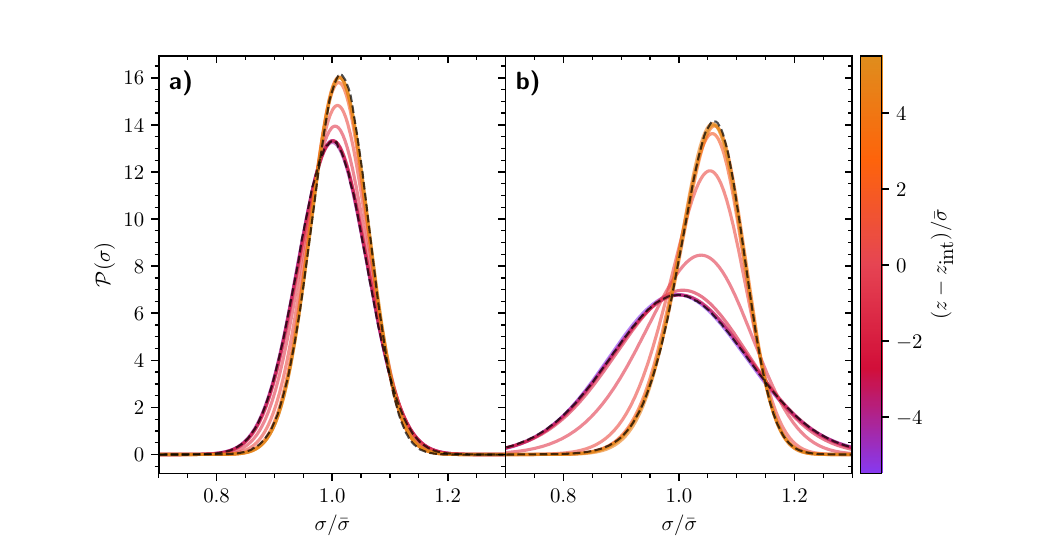}
    \caption{
        Scan of particle size distributions across the FCC$_a$ fluid-crystal interface for a system of 16000 particles where the fluid has polydispersity $p = 0.06$ (a) and $p = 0.12$ (b).
        The dashed black lines correspond to the respective size distributions of the reference fluid and crystal phases.
        Note that the range of sizes particles are allowed to explore in simulation far exceeds the range shown in this figure.
    }
    \label{fig:distribinterface}
\end{figure*}

Next, we examine how the distribution of particles varies in space as we move from one phase to the other through the interface. From the semi-grand simulations we performed at the coexistence point, we find the interface to be highly diffusive and very mobile.
Hence, to be able to characterize the interfacial structure, we shift  the position of all particles in each snapshot to correct for the motion of the interfaces. To this end, we monitor the mean displacement of the crystal in the elongated $z-$direction and shift the particles accordingly.
However, this trick does not prevent the crystal slab from growing or shrinking. As a second measure to correct for shifts in the positions of the two individual interfaces, we measure the ten Wolde $q_6$ bond orientational order parameters \cite{wolde1996simulation} as a function of $z$ and locate one interface at half the height of a fitted sigmoid function.

We measure the size distribution as a function of the distance to the interface along the long axis of the simulation box. Fits to these size distributions are shown in Fig. \ref{fig:distribinterface} for systems with polydispersity $p = 0.06$ and $p = 0.12$ in the FCC$_a$ orientation. These measurements corroborate what we previously reported in Fig. \ref{fig:phasediag}: as we progress from the fluid phase to the FCC crystal phase, the particle size distributions become increasingly more peaked and their means shift to larger sizes.
In other words, we see a crystal structure preferentially made of larger particles, and with less size diversity compared to what can be found in the fluid, consistent with what was predicted in Refs. \cite{coslovich2018local, fasolo2003equilibrium, fasolo2004fractionation, bolhuis1996monte,  kofke1999freezing, wilding2010phase, bolhuis1996numerical, sollich2010crystalline}.
As we show the same range for both scans in $z$ in Fig. \ref{fig:distribinterface}, we also note that the composition of the system can undergo drastic changes over a small distance.
Closer inspection of the size distributions in the crystal bulk and at the interface reveal that they deviate from the expected Gaussian behavior by being skewed towards the smaller sizes.

\begin{figure}
    \centering
    \includegraphics[width=\columnwidth]{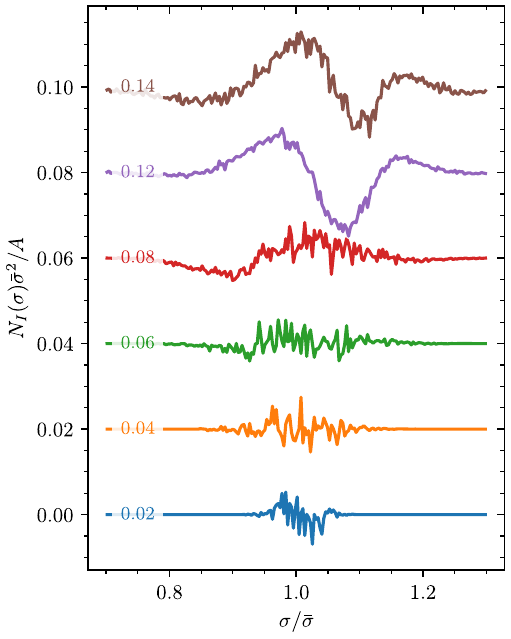}
    \caption{
        Surface excess amount of particles with size $\sigma_i$ for an equimolar Gibbs dividing surface in a system with 16384 particles at coexistence.
        Here the crystal lattice orientation is such that the square plane of the FCC crystal faces the fluid (FCC$_a$).
        Polydispersities are reported on each line, all lines are vertically shifted for readability.
    }
    \label{fig:excessa}
\end{figure}

Finally, our direct-coexistence simulations give us access to the interfacial excess absorption of particles of a given size $\sigma_i$.
This quantity reflects the excess amount (which can be either positive or negative) of a component -- here a particle of size $\sigma_i$ -- present in the system with respect to a reference system.
Following Gibbs formalism, we interpret our system as consisting of two bulk phases separated by a dividing surface which has no volume but can have particles absorbed to it. Hence, the total volume of the system is given by
\begin{equation}
    V = V_f + V_\chi,
\end{equation}
were the subscripts $f$ and $\chi$ refer respectively to the fluid and crystal phase. The total particle size distribution can be written as
\begin{equation}
    N(\sigma) = N_f(\sigma) + N_\chi(\sigma) + N_I(\sigma),
\end{equation}
where $N(\sigma)$ denotes the total number of particles of a given size $\sigma$, and the subscript $I$ refers to the interface. As our dividing surface, we choose the equimolar surface, which satisfies
\begin{equation}
    \int \mathrm{d}\sigma N_I(\sigma) = 0.
\end{equation}
We set up our coexistence simulations so that the global density of the box is precisely halfway in the coexistence region. For the equimolar surface, this implies that $V_F = V_X = V/2$.
The excess absorption at the interface $N_I(\sigma)$ is then given by
\begin{equation}
    N_I(\sigma) = N(\sigma) - \frac{V}{2}\left(\rho_f \mathcal{P}_f(\sigma) - \rho_\chi \mathcal{P}_\chi(\sigma) \right),
\end{equation}
where $\rho_f$ and $\rho_\chi$ are the coexistence densities of the fluid and crystal phase, respectively, and $\mathcal{P}_f$ and $\mathcal{P}_\chi$ are the corresponding particle size distributions. 
This quantity, normalized by the total interface area $A$, is reported in Fig. \ref{fig:excessa} for the range of polydispersities we investigated and for the FCC$_a$ interface.
We observe a significant change in the qualitative behavior of the surface excess which is strongly correlated with the crossing of the coexistence density lines we show in Fig. \ref{fig:phasediag}.
For the lower polydispersity regime ($p \leq 0.08$, leaving out $p = 0.02$), we observe a deficit of small particles and an excess of large particles at the interface, both of which are growing in amplitude with $p$.
For the higher polydispersity regime ($p \geq 0.12$), i.e. above the crossing of the density lines, this trend shifts entirely. We now see a net excess of both very small and very large particles, accompanied by a large deficit of particles with intermediate-to-large sizes.
Interestingly, at low polydispersities the picture is different for the FCC$_b$ interface (see SM), where we observe a deficit of the smaller particles and and excess of the larger ones from $p = 0.02$ and up.
We note that these measurements are likely to be sensitive to small inaccuracies in our determination of the coexistence conditions, as well as to long-time fluctuations in the amount of crystal present in the simulation box.
For reference, we show in the SM the composition of all investigated systems at coexistence as well as the composition of the reference fluid and crystal phases.

\section{Conclusion}
\label{section:conclusion}

In conclusion, we have introduced a novel method for locating phase boundaries in polydisperse systems, based on direct-coexistence simulations. Using this approach, we have mapped out the freezing line of polydisperse hard spheres with a Gaussian size distribution, finding good agreement with past predictions. 
Additionally, we have used the direct-coexistence simulations to probe the properties of the fluid-crystal interface.
We found that the surface stress increases as a function of polydispersity, so much as to become positive for $p \geq 0.06$, suggesting that particles at the interface adapt their sizes to reduce interfacial stress. This idea is corroborated by the observation of size-selectivity of excess particles associated with the interface.

While the discussion presented here is limited to coexistence between a Gaussian polydisperse fluid and a FCC crystal, a natural next step would be to use the same method to examine the stability of the more exotic phases that were seen to form spontaneously in highly polydisperse mixtures \cite{bommineni2019complex, lindquist2018communication, coslovich2018local}. Additionally, it would be interesting to examine how the choice of particle size distribution in the fluid phase impacts the emerging crystal phase.

We emphasize that the methodology presented here is not limited to hard spheres. Simulations in the canonical and semi-grand ensemble can be performed on systems interacting via any potential, and the only quantities we measure are particle forces (in the canonical ensemble) and pressure tensors (in the semi-grand ensemble), which are standard quantities of interest. The main caveat is that -- as seen in our case for polydispersity $p=0.10$, the direct-coexistence method struggles in systems where a stable interface is hard to form, which occurs when the density gap between the phases is very small. Additionally, in systems where dynamics are slow, equilibration of the coexisting system can take a long time, and hence be computationally expensive. Finally, we emphasize that the method proposed here is aimed at locating coexistences where the size distribution in one of the two phases (in this case the fluid) is known. This makes the method ideal for locating cloud and shadow curves, but unsuitable for finding the coexistence that would result from phase separation in the canonical ensemble, where the total size distribution (of both phases and the interface combined) would be fixed. For addressing the latter question, more complex simulation approaches or theoretical approximations are likely required, such as those developed in Refs. \onlinecite{sollich1998projected, wilding2009solid, sollich2010crystalline, sollich2011polydispersity}.
Despite these limitations, we believe that our approach provides an elegant and straightforward method for finding cloud and shadow curves in polydisperse systems.

\section{Supplementary Material}

Additional details for the generation of the deterministic particle size distribution as well as derivations for the effect of particle size on the free energy can be found in the supplementary material available online.
We also provide the derivation for the equations of motions in the semi-grand canonical ensemble, as well as implementation details for the EDMC approach. 
An overview of the effects of system size on the determination of coexistence conditions is also provided, along with additional details on the phase composition of the system at coexistence, and the excess surface absorption for the FCC$_b$ interface. 

\section{Acknowledgments}
We thank Marjolein de Jager for helpful discussions.
AC and FS acknowledge the use of the CERES high performance computer cluster at the Laboratoire de Physique des Solides of Université Paris-Saclay (Orsay, France).
AC and FS acknowledge funding from the Agence Nationale de la Recherche (ANR), grant ANR-21-CE30-0051. LF acknowledges funding from the Vidi research program with project number VI.VIDI.192.102 which is financed by the Dutch Research Council (NWO).

\section{Author Declarations}

\subsection{Conflict of Interest}

The authors have no conflicts to expose.

\section{Data Availability}
A version of the semi-grand canonical EDMC simulation code for polydisperse hard spheres is available online at \url{https://github.com/FSmallenburg/}.
The data that support the findings presented in this study are available in a data package on Zenodo \cite{castagnede2025polydatapackage}. The latter also includes additional surface forces data measured for a wide range of systems.

\bibliography{refs}

\end{document}